\documentclass[12 pt,twoside,aps,prd,amsmath,amssymb,
tightenlines,showpacs,showkeys,eqsecnum]{revtex4}
\usepackage{mathptmx}
\usepackage{bm}
\usepackage{graphicx}
\usepackage{hyperref}
\renewcommand{\cal}{\mathcal}

\newcommand {\cL}{\cal L}

\newcommand {\G}{\Gamma}
\newcommand {\bg}{\bar \gamma}
\newcommand {\bp}{\bar \psi}

\def \myfigures #1#2#3#4#5#6#7#8
{\begin{figure}[ht]
    \begin{center}
        \includegraphics[width=#2 \textwidth]{#1.eps}
        \hfill
        \includegraphics[width=#6 \textwidth]{#5.eps}
        \parbox[t]{#4\textwidth}{\caption {#3}\label{#1}}
        \hfill
        \parbox[t]{#8\textwidth}{\caption {#7}\label{#5}}
    \end{center}
\end{figure} }

\def\myfigure #1#2#3#4
{\begin{figure}[ht]\begin{center}
\includegraphics[width=#2 \textwidth]{#1.eps}
\parbox[t]{#4\textwidth}{\caption{#3}\label{#1}}
\end{center}\end{figure}}

\date{\today}
\begin{document}
\title{Nonlinear spinor field in Bianchi type-II spacetime}
\author{Bijan Saha}
\affiliation{Laboratory of Information Technologies\\
Joint Institute for Nuclear Research, Dubna\\
141980 Dubna, Moscow region, Russia} \email{bijan@jinr.ru}
\homepage{http://bijansaha.narod.ru}

\begin{abstract}

Within the scope of a Bianchi type-II (BII) cosmological model we
study the role of a nonlinear spinor field in the evolution of the
Universe. The system allows exact solutions only for some special
choice of spinor field nonlinearity.

\end{abstract}



\maketitle

\bigskip

\section{Introduction}

Recently, a number of authors showed the important role that spinor
fields play on the evolution of the Universe
\cite{henprd,sahagrg,sahajmp,sahaprd,SBprd04,BVI,ECAA06,sahaprd06,greene,kremer1,kremer2}.
In these papers the authors using the spinor field as the source of
gravitational one answered to some fundamental questions of modern
cosmology: (i) problem of initial singularity; (ii) problem of
isotropization and (iii) late time acceleration of the Universe. But
most of those works were done within the scope of either FRW or
Bianchi type-I cosmological models. Some works on Bianchi type-III,
V, VI$_0$ and VI were also done in recent time \cite{BVI,camb}. As
far as I know, there is still no work on Bianchi type-II model where
the author used spinor field as a source. In this report we plan to
fill up that gap.

\section{Basic equations}
We consider the simplest possible spinor field model within the
framework of a BI cosmological gravitational field given by the
Lagrangian density
\begin{equation}
{\cL} = \frac{R}{2\varkappa}  + \frac{i}{2} \biggl[\bp \gamma^{\mu}
\nabla_{\mu} \psi- \nabla_{\mu} \bar \psi \gamma^{\mu} \psi \biggr]
- m_{\rm sp}\bp \psi + F,\ \label{lag}
\end{equation}
where  $F(I,J), \quad I = S^2 = (\bp \psi)^2$ and $J = P^2 = (i \bp
\gamma^5 \psi)^2$ is the spinor field nonlinearity and $R$ is the
scalar curvature.

The gravitational field in our case is given by a Bianchi type-II
(BII) metric:
\begin{equation}
ds^{2} = dt^2 - a_1^2(dx + zdy)^2 - a_2^2 dy^2 - a_3^2 dz^2,
\label{BII}
\end{equation}
with $a_1,\,a_2,\,a_3$ being the functions of time only.

The spinor field equations corresponding to the metric \eqref{lag}
has the form
\begin{subequations}
\label{speq}
\begin{eqnarray}
i\gamma^\mu \nabla_\mu \psi - m_{\rm sp} \psi + F_S \psi + i F_P \gamma^5 \psi &=&0, \label{speq1} \\
i \nabla_\mu \bp \gamma^\mu +  m_{\rm sp} \bp - F_Ss \bp - i F_P \bp
\gamma^5 &=& 0, \label{speq2}
\end{eqnarray}
\end{subequations}
where $F_S = \frac{dF}{dS}$ and $F_P = \frac{dF}{dP}$. In
\eqref{lag}, and \eqref{speq} $\nabla_\mu$ is the covariant
derivative of spinor field:
\begin{equation}
\nabla_\mu \psi = \frac{\partial \psi}{\partial x^\mu} -\G_\mu \psi,
\quad \nabla_\mu \bp = \frac{\partial \bp}{\partial x^\mu} + \bp
\G_\mu, \label{covder}
\end{equation}
with $\G_\mu$ being the spinor affine connection. The spinor affine
connections for the metric \eqref{BII} has the form
\begin{subequations}
\label{SAC}
\begin{eqnarray}
\G_0 &=& 0, \nonumber\\
\G_1 &=&  \frac{1}{2} \dot a_1 \bg^1 \bg^0 + \frac{1}{4}
\frac{a_1^2}{a_2 a_3}\bg^2 \bg^3, \nonumber\\
\G_2 &=& \frac{1}{2} \dot a_2 \bg^2 \bg^0 + \frac{1}{2} z \dot a_1
\bg^1 \bg^0 + \frac{1}{4}\frac{a_1}{a_3}\bg^1 \bg^3 +  \frac{1}{4}
\frac{za_1^2}{a_2 a_3}\bg^2 \bg^3, \nonumber\\
\quad \G_3 &=& \frac{1}{2} \dot a_3 \bg^3 \bg^0 - \frac{1}{4}
\frac{a_1^2}{a_2} \bg^1 \bg^2.
\end{eqnarray}
\end{subequations}
It can be easily verified that
\begin{subequations}
\label{SAC1}
\begin{eqnarray}
\gamma^\mu \G_\mu &=& - \frac{1}{2} \Bigl(\frac{\dot a_1}{a_1} +
\frac{\dot a_2}{a_2} + \frac{\dot a_3}{a_3}\Bigr) \bg^0 -
\frac{1}{4} \frac{a_1^2}{a_2 a_3}\bg^1 \bg^2 \bg^3, \nonumber\\
\G_\mu \gamma^\mu  &=&  \frac{1}{2} \Bigl(\frac{\dot a_1}{a_1} +
\frac{\dot a_2}{a_2} + \frac{\dot a_3}{a_3}\Bigr) \bg^0 -
\frac{1}{4} \frac{a_1^2}{a_2 a_3}\bg^1 \bg^2 \bg^3, \nonumber
\end{eqnarray}
\end{subequations}
Defining
\begin{equation}
V = a_1 a_2 a_3, \label{Vdef}
\end{equation}
and taking into account that the spinor field is a function of $t$
only, one finds
\begin{subequations}
\label{speqn}
\begin{eqnarray}
\bg^0 \Bigl(\dot \psi + \frac{1}{2} \frac{\dot V}{V} \psi+
\frac{1}{4} \frac{a_1^2}{a_2 a_3} i \bg^5 \psi\Bigr)
+ i m_{\rm sp} \psi - i F_S \psi +  F_P \bg^5 \psi &=&0, \label{speq1n} \\
\Bigl(\dot \bp + \frac{1}{2} \frac{\dot V}{V} \bp+ \frac{1}{4}
\frac{a_1^2}{a_2 a_3} i \bp \bg^5\Bigr) \bg^0 - i  m_{\rm sp} \bp +
i F_S \bp -  F_P \bp \bg^5 &=& 0, \label{speq2n}
\end{eqnarray}
\end{subequations}
From \eqref{speqn} one finds
\begin{subequations}
\label{speqinv}
\begin{eqnarray}
\dot S +  \frac{\dot V}{V} S + \frac{1}{2} \frac{a_1^2}{a_2 a_3} P -
2 F_P A^0 &=&0, \label{speq1nS} \\
\dot P +  \frac{\dot V}{V} P - \frac{1}{2} \frac{a_1^2}{a_2 a_3} S -
2  m_{\rm sp} A^0 + 2 F_S A^0 &=& 0, \label{speq2nP}\\
\dot A^0 +  \frac{\dot V}{V} A^0 + 2 m_{\rm sp} P - 2 F_S P + 2 F_P
S &=& 0, \label{speq3nA0}
\end{eqnarray}
\end{subequations}
where, $A^0 = \bp \bg^5 \bg^0 \psi$. From \eqref{speqinv} one finds
\begin{equation}
V^2\bigl(S^2 + P^2 + A^{02}\bigr) = {\rm Const.} \label{inv3}
\end{equation}
Note that, from \eqref{speq1nS} and \eqref{speq2nP} one finds
\begin{equation}
\frac{1}{2} \frac{\partial}{\partial t} \bigl(S^2 + P^2\bigr) +
\frac{\dot V}{V} \bigl(S^2 + P^2\bigr) - 2 \bigl(F_P S - F_S
 P\bigr) A^0 = 0. \label{invsp}
\end{equation}
As one sees, the assumption
\begin{equation}
F_P S - F_S P = 0, \label{assump}
\end{equation}
leads to
\begin{equation}
V^2 \bigl(S^2 + P^2\bigr) = C_0^2, \quad C_0^2 = {\rm Const.}
\label{inv2}
\end{equation}
It can be easily verified that the relation \eqref{assump} holds, if
one assumes that $F = F(S^2 + P^2)$.

Let us now write the components of energy momentum tensor for the
spinor field. In the case considered, one finds

\begin{equation}
T_0^0 = m_{\rm sp} S - F, \quad T_1^1 = T_2^2 = T_3^3 = S F_S + P
F_P - F. \label{tem}
\end{equation}
Let us now write the Einstein field equations corresponding to BII
metric \eqref{BII}. As it was shown in a recent paper
\cite{sahacejp2011}, thanks to $T_1^1 = T_2^2 = T_3^3$ the
off-diagonal component of the Einstein equation can be overlooked.
As a result we now have the following system:
\begin{subequations}
\label{einBII}
\begin{eqnarray}
\frac{\ddot a_2}{a_2} +\frac{\ddot a_3}{a_3} +\frac{\dot
a_2}{a_2}\frac{\dot a_3}{a_3} - \frac{3}{4}\frac{a_1^2}{a_2^2 a_3^2}
&=&  \kappa \bigl(S F_S + P F_P - F\bigr), \label{11BII}\\
\frac{\ddot a_3}{a_3} +\frac{\ddot a_1}{a_1} +\frac{\dot
a_3}{a_3}\frac{\dot a_1}{a_1} + \frac{1}{4}\frac{a_1^2}{a_2^2 a_3^2}
&=&  \kappa \bigl(S F_S + P F_P - F\bigr), \label{22BII} \\
\frac{\ddot a_1}{a_1} +\frac{\ddot a_2}{a_2} +\frac{\dot
a_1}{a_1}\frac{\dot a_2}{a_2} + \frac{1}{4}\frac{a_1^2}{a_2^2 a_3^2}
&=& \kappa \bigl(S F_S + P F_P - F\bigr), \label{33BII}\\
\frac{\dot a_1}{a_1}\frac{\dot a_2}{a_2} +\frac{\dot a_2}{a_2}
\frac{\dot a_3}{a_3} + \frac{\dot a_3}{a_3}\frac{\dot a_1}{a_1} -
\frac{1}{4}\frac{a_1^2}{a_2^2 a_3^2} &=& \kappa\bigl(m_{\rm sp} S - F\bigr). \label{00BII}\\
\end{eqnarray}
\end{subequations}
Subtracting \eqref{22BII} from \eqref{33BII} one finds
\begin{equation}
\frac{\ddot a_2}{a_2} - \frac{\ddot a_3}{a_3} + \frac{\dot
a_1}{a_1}\Bigl(\frac{\dot a_2}{a_2} - \frac{\dot a_3}{a_3} \Bigr) =
0, \label{a23}
\end{equation}
which yields
\begin{equation}
\frac{a_3}{a_2} = D_2 \exp{[X_2 \int V^{-1} dt]}, \label{a23n}
\end{equation}
with $D_2$ and $X_2$ being some arbitrary constants.

Summation of \eqref{11BII}, \eqref{22BII}, \eqref{33BII} and 3 times
\eqref{00BII} gives the equation for $V$:
\begin{equation}
2 \frac{\ddot V}{V} =  \frac{a_1^2}{a_2^2 a_3^2} + \kappa [m_{\rm
sp} S + 3 (S F_S + P F_P - 2F)]. \label{eqV}
\end{equation}

The right hand side of \eqref{eqV} explicitly depends on $a_1, a_2 $
and $a_3$. We need some additional conditions to overcome it.
Following many authors we assume the expansion $\vartheta$ is
proportional to any of the components (say $\sigma_1^1$) of the
shear tensor $\sigma$. We will choose a comoving frame of reference
so that $u_\mu = (1,\,0,\,0,\,0)$ and $u_\mu u^\mu = 1.$  In this
case we find
\begin{equation}
\vartheta =  \Gamma^\mu_{\mu 0} = \frac{\dot{a}}{a} +
\frac{\dot{b}}{b} + \frac{\dot{c}}{c} = \frac{\dot{V}}{V},
\label{expanBII}
\end{equation}
and
\begin{equation}
\sigma_1^1 = \frac{\dot a}{a} - \frac{1}{3} \vartheta.
\label{shearmix11II}
\end{equation}
The proportionality condition
\begin{equation}
\sigma_1^1 = q_1 \vartheta, \qquad q_1 = {\rm const.}
\label{propcon}
\end{equation}
leads to
\begin{equation}
a_1 = (a_2 a_3)^{(1 + 3q_1)/(2-3q_1)}. \label{propcon1}
\end{equation}
On account of \eqref{propcon1}, \eqref{Vdef} and \eqref{a23n} one
finds
\begin{subequations}
\begin{eqnarray}
a_1 &=& V^{(1 + 3q_1)/3}, \label{a1}\\
a_2 &=& (1/\sqrt{D_2}) V^{(2-3q_1)/6} e^{-\frac{X_2}{2} \int
\frac{dt}{V}}, \label{a2}\\
a_3 &=& \sqrt{D_2})V^{(2-3q_1)/6} e^{\frac{X_2}{2} \int
\frac{dt}{V}}. \label{a3}
\end{eqnarray}
\end{subequations}

The Eq. \eqref{eqV} now can be written as

\begin{equation}
2 \ddot V =  V^{(12q_1+1)/3} + \kappa [m_{\rm sp} S + 3 (S F_S + P
F_P - 2F)]V. \label{eqVn}
\end{equation}
To this end we assume that the spinor field be a massless one and
the spinor field nonlinearity is given by $F = F(K)$ with $K = S^2 +
P^2$. In this case $F_S = 2 S F_K$ and $F_P = 2 P F_K$, hence $S F_S
+ P F_P  = 2 (S^2 + P^2) F_K = 2 K F_K$. The Eq. \eqref{eqVn} then
reads
\begin{equation}
2 \ddot V =  V^{(12q_1+1)/3} + 6 \kappa [K F_K - F]V. \label{eqVnK}
\end{equation}
Let us now choose the spinor field nonlinearity in some concrete
form. We will consider the case when $F$ is a power law of $K$,
namely, $F = K^n$. Taking into account that $K = S^2 + P^2 =
C_0^2/V^2$ we rewrite \eqref{eqVnK} as
\begin{equation}
2 \ddot V =  V^{(12q_1+1)/3} + 6 \kappa (n-1) C_0^{2n} V^{1-2n},
\label{eqVnK1}
\end{equation}
with the solution in quadrature
\begin{equation}
\int \frac{dV}{\sqrt{[3/(12q_1 + 4)]V^{(12q_1+4)/3} - 3 \kappa
C_0^{2n} V^{2-2n} + C_1}} = t + t_0, \label{quad}
\end{equation}
with $C_1$ and $t_0$ being integration constants. As one sees, in
the model considered, the Heisenberg-Ivanenko type nonlinearity with
$n = 1$ has no influence on the evolution of the Universe.

\section{Conclusion}

Within the scope of a Bianchi type-II cosmological model the role of
a nonlinear spinor field on the evolution of the Universe is
studied. It is shown that the model allows exact solutions only for
some special choice of nonlinearity. In the case considered the
isotropization process of the initially anisotropic spacetime does
not take place.

\end{document}